\documentclass[fdp,a4paper,fleqn]{w-art}
\usepackage{times,cite,w-thm}
\theoremstyle{plain}

\theoremstyle{definition}

\usepackage[]{graphicx}
\begin{document}
\DOIsuffix{theDOIsuffix} 
\Volume{XX} \Month{XX} \Year{2012} \pagespan{1}{}


\title[Semiconductor `billiards': boundaries versus disorder]{Is it the boundaries or disorder that dominates electron transport in semiconductor `billiards'?}

\author[A.P.~Micolich]{A.P. Micolich\inst{1}
  \footnote{Corresponding author\quad E-mail:~\textsf{adam.micolich@nanoelectronics.physics.unsw.edu.au}}}
\author[A.M.~See]{A.M. See\inst{1}}
\author[B.C.~Scannell]{B.C. Scannell\inst{2}}
\author[C.A.~Marlow]{C.A. Marlow\inst{2}}
\author[T.P.~Martin]{T.P. Martin\inst{1,2}
  \footnote{Current address: Acoustics Division, Naval Research Laboratory, Washington, DC 20375, U.S.A.}}
\author[I.~Pilgrim]{I. Pilgrim\inst{2}}
\author[A.R.~Hamilton]{A.R. Hamilton\inst{1}}
\author[H.~Linke]{H. Linke\inst{3}}
\author[R.P.~Taylor]{R.P. Taylor\inst{2}}

\address[\inst{1}]{School of Physics, The University of New South Wales, Sydney NSW 2052, Australia}
\address[\inst{2}]{Materials Science Institute, Physics Department, University of Oregon, Eugene OR 97403-1274, U.S.A.}
\address[\inst{3}]{Solid State Physics and nmC@LU, Lund University, Box 118, SE-221 00, Lund, Sweden}

\begin{abstract}
Semiconductor billiards are often considered as ideal systems for
studying dynamical chaos in the quantum mechanical limit. In the
traditional picture, once the electron's mean free path, as
determined by the mobility, becomes larger than the device, disorder
is negligible and electron trajectories are shaped by specular
reflection from the billiard walls alone. Experimental insight into
the electron dynamics is normally obtained by magnetoconductance
measurements. A number of recent experimental studies have shown
these measurements to be largely independent of the billiards exact
shape, and highly dependent on sample-to-sample variations in
disorder. In this paper, we discuss these more recent findings
within the full historical context of work on semiconductor
billiards, and offer strong evidence that small-angle scattering at
the sub-$100$~nm length-scale dominates transport in these devices,
with important implications for the role these devices can play for
experimental tests of ideas in quantum chaos.
\end{abstract}

\maketitle
\renewcommand{\leftmark} {A.P. Micolich {\it et al.}: Semiconductor billiards: boundaries versus disorder}

\tableofcontents

\section{Introduction}

Quantum mechanics and chaos theory are two great discoveries of the
$20^{th}$ century. Combining them leads to highly interesting and
important questions; for example, how does the uncertainty
principle's inherent `blurriness' affect the sensitivity to initial
conditions characteristic of chaotic systems? Quantum chaos rose to
popular attention in the 1970s, first in the context of electron
dynamics in atoms~\cite{GutzwillerJMP71, BerryPRSLA76}, and
subsequently for model two-dimensional (2D) dynamical systems known
as `billiards'~\cite{McDonaldPRL79, HellerPRL84}. From a classical
perspective, a billiard contains a particle traveling in straight
lines between specular reflections at a shaped boundary wall. A
variety of billiard shapes have been studied ranging from simple
geometries (e.g., circle or square) to more complex geometries
(e.g., stadia and `lemon' billiards~\cite{HellerPT93}) and
geometries with multiple topologically-independent walls (e.g., the
Sinai billiard~\cite{SinaiRMS70}). Billiards have since become an
archetypal system for studies of quantum
chaos~\cite{GutzwillerBook90, StockmannBook99}.

The quest for experimental realizations of quantum billiards
naturally followed, with microwave billiards formed in open
resonators~\cite{StockmannPRL90, SridharPRL91, SteinPRL92} and
superconducting slabs~\cite{GrafPRL92} leading the way, closely
followed by semiconductor microstructures~\cite{MarcusPRL92,
ChangPRL94, BerryPRB94}. In the years that followed, quantum
billiards have also been realized in quantum `corrals' defined by
scanning tunneling microscopy on metal surfaces~\cite{CrommieSci94,
HellerNat94}, in acoustic systems~\cite{EllegaardPRL95,
SchaadtPRE03}, atom-optics systems~\cite{MilnerPRL01,
FriedmanPRL01}, and various optical systems including micro-cavity
lasers~\cite{GmachlSci98} and double-clad optical
fibers~\cite{DoyaOL01, DoyaPRL02}.

An important consideration in comparing measurements to theoretical
expectations is to adequately account for practical aspects of the
experimental implementation that affect the correspondence. For
semiconductor billiards, two important aspects are the electrostatic
nature of confinement and impurities in the
semiconductor~\cite{BerggrenChaos96}. These aspects, impurity
scattering in particular, were considered in initial
studies~\cite{MarcusPRL92, MarcusChaos93, LinChaos93, ChangPRL94}
and suggested to not obscure the essential physics. This has
resulted in a widely-held perception that semiconductor
microstructures are an ideal test-bed for dynamical quantum chaos;
for example, {\it ``... it is important to understand the
consequences of non-diffusive electron dynamics on the electronic
conductance or other transport properties. This question has been
studied in much detail for semiconductor nanostructures in which the
motion of electrons is ballistic rather than diffusive. In such
systems, disorder is negligible, and, consequently, all transport
properties are determined by the shape of the sample, as in a
billiard model."}~\cite{PrustyPRL06}. However, a range of
experiments following these initial studies have gradually shown
that impurity scattering not only cannot be ignored, but may
actually dominate over boundary-dependent `ballistic' transport in
semiconductor billiards.

Here, we provide a short, focussed review of studies of electron
transport in semiconductor microstructures, and offer evidence that
these devices are not the ideal embodiment of quantum dynamical
billiards, as is often thought based on the pioneering papers on
this topic~\cite{JalabertPRL90, MarcusPRL92, BarangerPRL93,
ChangPRL94, JensenNat95}. Our aim is to show that the true picture
of transport in these devices is much more complex than initially
anticipated.

\section{Semiconductor quantum dots - The fundamentals}
\subsection{Making quantum dots}

The underpinning structure for most studies of semiconductor quantum
dots and billiards is the modulation-doped AlGaAs/GaAs
heterostructure~\cite{DingleAPL78}. It typically consists of a GaAs
substrate supporting an undoped AlGaAs spacer layer, a Si-doped
AlGaAs modulation-doping layer, and an undoped GaAs cap layer
(Fig.~1(b)) grown using molecular beam epitaxy
(MBE)~\cite{ChoAPL71}. The narrow, triangular potential well at the
AlGaAs/GaAs interface forms self-consistently through the
band-bending induced by ionization of the Si dopants (Fig.~1(c)).
The well width is comparable to the electron Fermi wavelength
($\lambda_{F} \sim 50$~nm), strongly confining the electrons in the
growth direction to form a two-dimensional electron gas (2DEG).
Further confinement of the 2DEG is typically achieved using a set of
metal surface electrodes (gates) defined using electron beam
lithography techniques~\cite{ThorntonPRL86, TaylorNano94}
(Fig.~1(a)). Application of a negative bias $V_{g}$ to the gates
electrostatically depletes the 2DEG directly underneath,
transferring the gate pattern into the 2DEG as shown in
Fig.~1(b).\footnote{A common alternative is to selectively etch away
the doping, which also gives patterned depletion regions in the
2DEG.} The result is a quasi-zero-dimensional quantum dot connected
by small one-dimensional (1D) entrance and exit ports to 2DEG source
and drain reservoirs. These are connected to the external circuit
via diffused AuGeNi ohmic contacts.

\begin{figure}
\begin{center}
\includegraphics[width=0.8\linewidth]{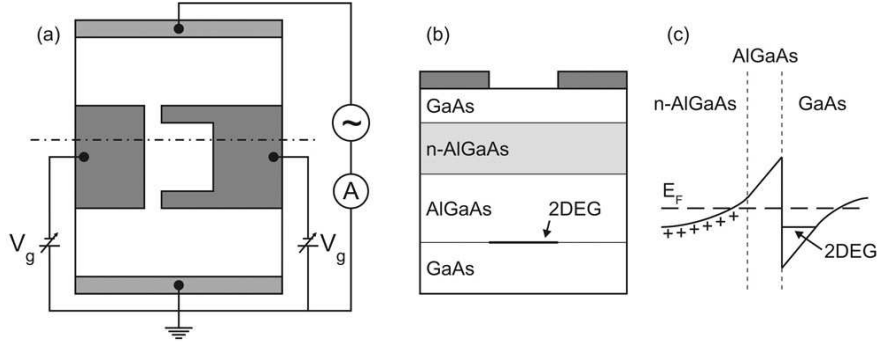}
\caption{(a) Top view of a semiconductor billiard including basic
measurement circuit. The billiard is defined by a pair of metal
surface electrodes (gates) with a negative d.c. bias $V_{g}$
applied, and connected to the measurement circuit via 2DEG source
and drain reservoirs, each featuring an AuGeNi ohmic contact. The
dot-dashed line indicates the cross-section presented in (b). (b)
heterostructure cross-section illustrating electrostatic depletion
due to the negatively-biased gates. (c) Conduction band diagram
illustrating 2DEG formation at the AlGaAs/GaAs interface. The dashed
line indicates the electron Fermi energy $E_{F}$ and $+$ the ionized
Si dopants in the n-AlGaAs layer.}
\end{center}
\end{figure}

\subsection{Length scales and transport regimes}

A hallmark of MBE-grown AlGaAs/GaAs heterostructures is an extremely
high electrical mobility, typically $\mu \sim 10^{5} -
10^{7}$~cm$^{2}$/Vs, corresponding to an electron mean free path
$\ell \sim 0.5 - 50~\mu$m at a typical 2DEG electron density $n \sim
10^{11}$~cm$^{-2}$. The traditional paradigm of transport regimes in
mesoscopic devices~\cite{BeenakkerSSP91} depends on the relationship
between $\ell$ and the device's length $L$ and width $W$
(Fig.~2(a-c)). In the diffusive regime, $\ell < L,W$, the electron's
trajectory is primarily determined by the impurity configuration
(Fig.~2(a)). In the ballistic regime, $\ell > L,W$, there are no
impurities within the device on average, and an electron follows
straight line paths between specular reflections at the device
boundary (Fig.~2(c)). In the quasi-ballistic regime, $\ell \sim L,W$
or $W < \ell < L$, a mixture of impurity and boundary scattering
determines the electron's trajectory (Fig.~2(b)). The quantum dots
used as billiards typically have $L,W < 1~\mu$m and $\ell
> 2~\mu$m.

\begin{figure}[b]
\includegraphics[width=\linewidth]{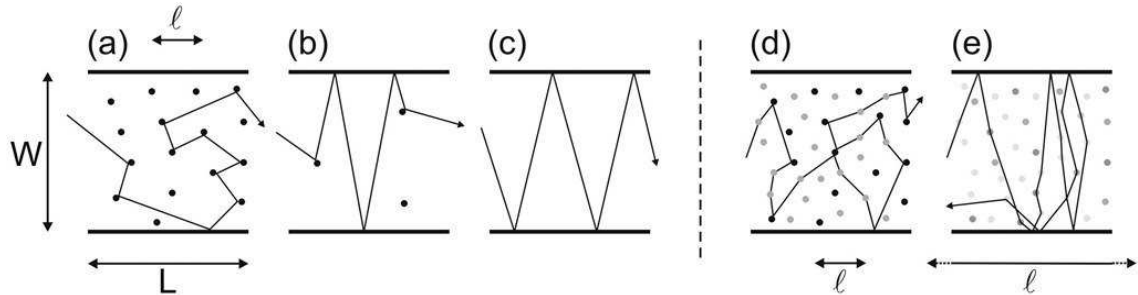}
\caption{(a-c) The traditional paradigm of transport regimes in
mesoscopic devices~\cite{BeenakkerSSP91}: (a) diffusive $\ell <
L,W$, (b) quasi-ballistic $\ell \sim L,W$, and (c) ballistic $\ell
> L,W$. $L$ and $W$ are the device/channel length and width
and $\ell$ is the electron mean free path determined from the
electrical mobility $\mu$. (d/e) Schematics for a modified picture
of the (d) diffusive and (e) ballistic regimes accounting for
small-angle ionized impurity scattering, as discussed in $\S4.2$.
The lighter dots in (d/e) correspond to small-angle scatterers, with
darker shades representing a greater propensity towards larger
scattering angles.}
\end{figure}

For later discussion, it is worth bearing in mind that this is a
somewhat simplistic and incomplete picture of disorder in these
devices, because $\mu$ is heavily weighted towards large-angle
scattering events~\cite{ColeridgePRB91}. As a result, small-angle
scattering is often under-appreciated; we will return to this in
$\S4.2$.

\subsection{Quantum interference and conductance fluctuations}

The Aharonov-Bohm (AB) effect~\cite{AharonovPR59} demonstrates the
influence of electromagnetic potentials on quantum interference. The
original concept involved an infinite length solenoid, with an
electron beam split and recombined in the plane perpendicular to the
solenoid axis to form an interferometer `loop'. For an infinitely
long solenoid, the field $B$ experienced by the electrons is zero;
the AB effect is caused by the magnetic vector potential ${\bf A}$,
such that electrons traversing opposite arms of the loop accumulate
phase terms of opposing sign $\pm (e/\hbar) \int{\bf A}\cdot dl$,
where $dl$ is an element of path length. The two electron beams
interfere when recombined, giving a transmitted beam intensity that
is an oscillatory function of the flux $\phi$ enclosed by the loop
(and in this case, restricted entirely to the solenoid's interior).
Early electron microscope experiments demonstrated flux-induced
electron interference~\cite{ChambersPRL60}, but it took another 25
years to demonstrate interference occurring via the vector potential
${\bf A}$ alone~\cite{TonomuraPRL86}.\footnote{Superconducting films
were used to screen the electron beams from any magnetic field $B$
generated whilst establishing ${\bf A}$.}

\begin{figure}
\begin{center}
\includegraphics[width=0.7\linewidth]{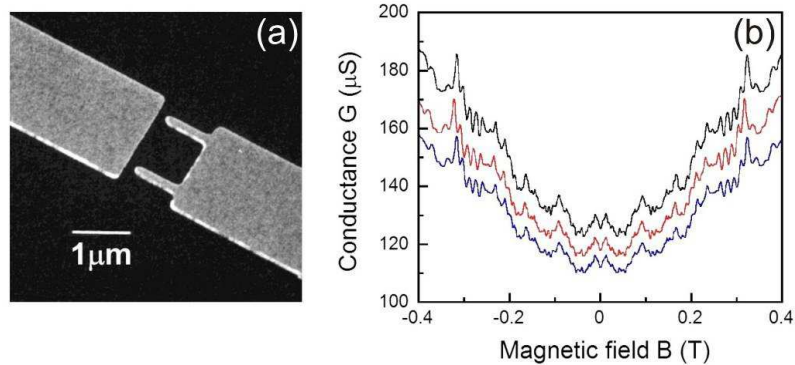}
\caption{(a) Scanning electron micrograph of a typical quantum dot
device with Ti/Au gates (grey) on the GaAs surface (black). The $1
\times 1~\mu$m quantum dot is connected to 2DEG source and drain
reservoirs via 1D quantum point contact (QPC) entrance and exit
ports. (b) Conductance $G$ versus magnetic field $B$ perpendicular
to the 2DEG from the device in (a). Three traces are shown; the
upper and middle traces are subsequent sweeps of $B$, the lower
trace was obtained $10$~hours later, on returning the device to the
same $V_{g}$ used for the first two traces after other
measurements.}
\end{center}
\end{figure}

The Aharonov-Bohm effect lays at the heart of a range of magnetic
field induced fluctuations in the electrical conductance of
mesoscopic devices. At its simplest, it produces a sinusoidal
fluctuation with a period $\Delta B = h/eA$ in the conductance $G$
of nanoscale ring structures enclosing an area $A$; observed in both
gold rings~\cite{WebbPRL85} and etched ring-shaped structures in an
AlGaAs/GaAs heterostructure~\cite{TimpPRL87}. These experiments are
performed at a temperature $T < 1$~K using a $^{3}$He cryostat or a
$^{3}$He/$^{4}$He dilution refrigerator to ensure electron phase
coherence. The conductance fluctuations develop richer structure
when the interference involves a larger number of possible electron
paths through the device. A classic example is the $1~\mu$m square
quantum dot in Fig.~3(a), which is connected to 2DEG source and
drain reservoirs via a pair of Quantum Point Contact (QPC) entrance
and exit ports. The QPC strongly diffracts the incoming electron
wave~\cite{AkisJJAP97, BirdCSF97}, allowing a range of trajectories
linking the entrance and exit QPCs to interfere and contribute to
$G$. In a semiclassical picture, each trajectory pair encloses a
different area $\theta$, giving a slightly different period of AB
oscillation; these combine to yield reproducible multi-spectral
magnetoconductance $G(B)$ fluctuations (Fig.~3(b)). These
fluctuations are not noise; as the lower trace in Fig.~3(b)
demonstrates, identical fluctuations are obtained if a device
returned to a given gate bias configuration after measurements at
other biases, providing the temperature remains low throughout (see
$\S4.2$). The link between the spectral content of the
magnetoconductance fluctuations (MCF) and the distribution of loop
areas $N(\theta)$~\cite{TaylorSurfSci88} provides a
`magnetofingerprint' of the electron trajectories within the
dot~\cite{FengPRL86, LeePRB87}. While the underlying mechanism for
these fluctuations is the same as that driving the well-known
Universal Conductance Fluctuations (UCF)~\cite{LeePRL85,
AltshulerJETP85, LiciniPRL85, KaplanPRL86, SkocpolPRL86, LeePRB87,
SkocpolPS87}, the amplitude of the fluctuations for dots is not
universal~\cite{MarcusPRL92}; this is due to the lack of ensemble
averaging that occurs in diffusive systems~\cite{LeePRL85,
AltshulerJETP85, SkocpolPRL86, SkocpolPS87}. Another commonly held
difference between UCF in metals or MOSFETs and MCF in quantum dots
is that whereas in the former transport is diffusive, in the latter
the transport is assumed to be ballistic and hence determined by the
geometry of the dot walls. We will return to examine this assumption
in $\S4.2$ in light of recent experiments~\cite{ScannellArXiv11,
SeePRL12} that question the way we think about semiconductor
billiards as systems for studying quantum chaos.

\section{Early pictures of quantum chaos in semiconductor billiards}

The use of ballistic microstructures for studies of quantum
chaos~\cite{BeenakkerPRL89} evolved from experimental studies of
ballistic junction devices~\cite{FordPRL89, ChangPRL89} in the late
1980s. In particular, a seminal theoretical paper by Jalabert {\it
et al}~\cite{JalabertPRL90} proposed that knowledge of the chaotic
classical scattering dynamics could be used to predict
quantitatively measurable properties of ballistic microstructures.
This relies on a universal property of chaotic dynamics in open
scattering systems whereby a particle's probability of escape is
exponential in time~\cite{TelPRA87}, giving measurable correlations
in the energy/frequency spectrum~\cite{BlumelPRL90, DoronPRL90}.
Jalabert {\it et al} extended this property to ballistic
microstructures, translating the escape time distribution into a
distribution of accumulated trajectory areas, which, using
semiclassical arguments, produces MCF with particular statistical
properties via the AB effect. In particular, the autocorrelation
function $C(\Delta B) = <\delta g(B)\delta g(B+\Delta B)>$ for the
MCF should take the form:

\begin{equation}
  C(\Delta B) = C(0)/[1+(\Delta B/\alpha\phi_{0})^{2}]^{2}
\end{equation}

\noindent where $\delta g(B)$ is the fluctuation from mean
conductance. Here $\phi_{0} = h/e$ and $\alpha$ is the exponent for
the distribution $N(\theta) \propto \exp (-2\pi\alpha|\theta|)$ of
areas $\theta$ enclosed by various possible
trajectories~\cite{JalabertPRL90, MarcusPRL92}.

Marcus {\it et al}~\cite{MarcusPRL92} reported the first experiment
devoted specifically to chaotic scattering in ballistic
microstructures, measuring two separate chips, each containing two
quantum dots, one with a circular geometry and one with a stadium
shaped geometry. The study had two intentions: to confirm the
prediction by Jalabert {\it et al}~\cite{JalabertPRL90} for chaotic
scattering systems, and to confirm predictions of quantitative
statistical differences in the fluctuations between geometries
supporting chaotic (stadium) and non-chaotic (circle)
transport~\cite{OakeshottSM92, NakamuraJPSJ92}. Comparison with
theoretical expectations in Ref.~\cite{JalabertPRL90} was made via
the Fourier transform of Eq.~1, which gives the MCF power
spectrum~\cite{MarcusPRL92}:

\begin{equation}
  S(f) = S(0)[1+(2\pi\alpha\phi_{0})f]e^{-2\pi\alpha\phi_{0}f}
\end{equation}

\noindent Fits of Eq.~2 for the stadium geometry gave
$\alpha\phi_{0} \sim 3.6$~mT, corresponding to an $\alpha^{-1} \sim
2.7$ times the dot area $A$, consistent with expectations from
Ref.~\cite{JalabertPRL90}. The spectral power $S(f)$ for the circle
was similar at $f < 250$~T$^{-1}$ but higher at $250 < f <
1000$~T$^{-1}$, consistent with predictions of enhanced higher
frequency content, corresponding to a larger number of large-area
trajectories in non-chaotic geometries~\cite{OakeshottSM92,
NakamuraJPSJ92}. Note that the power spectra used for comparison to
Ref.~\cite{JalabertPRL90} were heavily averaged ($15$ fast Fourier
transform (FFT) spectra from half-overlapping $256$-point
($18.6$~mT) wide windows). Single FFT spectra for $1024$ point
blocks spanning $-70 < B < 70$~mT revealed structure at low
frequencies, consistent with structure in the large $\Delta B$ tails
of the $C(\Delta B)$ versus $\Delta B$ data presented in
Ref.~\cite{MarcusPRL92}. Such low frequency spectral peaks are an
expected signature of wavefunction scarring by periodic
orbits~\cite{HellerPRL84, WilkinsonNat96, AkisJJAP97}, and have been
extensively studied in subsequent calculations~\cite{AkisJJAP97,
BirdCSF97, AkisPRL97}, transport studies~\cite{BirdPRL99} and
scanning-gate microscopy (SGM) studies~\cite{CrookPRL03, BurkePRL10,
AokiPRL12}. However, of particular note for the discussion in
$\S4.2$ is a comparison between the power spectra for nominally
identical geometries (i.e., the stadium on device 1 with the stadium
on device 2 or the circle on device 1 with the circle on device 2)
in Fig.~3 of Ref.~\cite{MarcusPRL92}. If the dominant factor in
determining the trajectory distribution is device geometry, then the
spectra for identical geometries on different chips should match, or
at least be very similar; they instead differ significantly,
indicating that dot shape alone cannot account for the electron
dynamics.

Subsequent theoretical work by Baranger {\it et
al}~\cite{BarangerPRL93} proposed the alternative possibility of
distinguishing between geometries supporting chaotic and regular
(i.e., non-chaotic) dynamics using the `ballistic'
analog~\cite{MarcusPRL92} of another important interference effect
in diffusive 2D systems known as weak localization
(WL)~\cite{BergmannPR84}. Weak localization involves interference
between a trajectory that returns to its point of origin via a set
of impurity scattering events, and its time-reversed counterpart.
Application of an increasing perpendicular magnetic field $B$
progressively breaks time-reversal symmetry for shorter length
paths, leading to a positive magnetoconductivity correction (minima
in $G(B)$ at $B = 0$), and hence a peak in the
magnetoresistance.~\footnote{n.b. alternative explanations for the
magnetoconductivity about $B = 0$ in quantum dots have also been
subsequently proposed~\cite{AkisPRB99, AkisJPCM99, BirdPRB99}.} The
proposal by Baranger {\it et al}~\cite{BarangerPRL93} is that the
different dynamics produce different WL peak lineshapes, which
relies on the distribution of enclosed loop areas $N(\theta)$
differing between chaotic and regular billiards. Under a
semiclassical framework, the average quantum correction to the
reflection coefficient $R_{D}$ for chaotic billiards takes the form:

\begin{equation}
  \langle\delta R_{D}(B)\rangle = \textsl{R}/[1 + (2B/\alpha\phi_{0})^{2}]
\end{equation}

\noindent where $\textsl{R} = \langle\delta R_{D}(B =
0)\rangle$~\cite{BarangerChaos93}, and for regular billiards:

\begin{equation}
  \langle\delta R_{D}(B)\rangle \propto |B|
\end{equation}

\noindent for small $B$~\cite{BarangerPRL93}. It is important that
the average above is taken over an infinite window in $k$ to ensure
that the contributions to Eq.~3/4 only include the interference of
paths that are precise time-reversed pairs. This averaging has
important experimental implications -- the geometry effect should
not be directly observable in a single magnetoresistance $R(B)$
trace, but only when sufficient averaging is applied to eliminate
the conductance fluctuations~\cite{BarangerChaos93}. Indeed, it is
reported in Ref.~\cite{BarangerChaos93} that calculations of
unaveraged $R(B)$ traces yield a {\it minimum} at $B = 0$ in $\sim
33\%$ of cases, in contrast to the expected $B = 0$ maxima. This is
borne out experimentally; e.g., in studies of dots featuring an
additional `shape-distorting' gate by Chan {\it et
al}~\cite{ChanPRL95} averaging was needed to obtain a clear
Lorentzian $R(B)$ peak and $R(B)$ minima are occasionally
observed~\cite{MarcusPC}, in studies of a square billiard by Bird
{\it et al}~\cite{BirdPRB95} where $R(B)$ peak lineshape (Lorentzian
vs linear) was sensitive to gate voltage, and in studies by Keller
{\it et al}~\cite{KellerPRB96} where $G(B)$ maxima and minima were
obtained at different densities (i.e., Fermi energy) while the
ensemble-average $\langle G(B)\rangle$ gave the expected minima at
$B = 0$.

Measurements by Chang {\it et al}~\cite{ChangPRL94} provided
evidence in support of the prediction by Baranger {\it et al} above;
reporting a Lorentzian $R(B = 0)$ peak for a stadium geometry
(chaotic) and a linear $R(B = 0)$ peak for a circular geometry
(non-chaotic). It is very important to note that these studies were
not performed in single dots, but in devices consisting of 48 dots
with nominally-identical geometry arranged in a $6 \times 8$ array.
This is vital in terms of the averaging issues discussed in
Ref.~\cite{BarangerChaos93}, but may also have important
implications in terms of small-angle scattering, as we will discuss
in $\S5$.

This brings us to an important turning point in terms of our
understanding of ballistic transport and quantum chaos in
semiconductor billiards and quantum dots. On the basis of the work
discussed above it is tantalizing and easy to assume a great
simplicity in these devices: images of electrons traveling along
straight line trajectories between specular reflections from walls
with well defined geometrical shapes, clear divisions of geometries
into chaotic and non-chaotic, which in turn give clear and distinct
experimental signatures, etc. see, e.g., Ref.~\cite{JensenNat95}. As
the following discussion will show, in reality, the picture of
electron dynamics in semiconductor billiards is more complex. The
practicalities involved in the experimental realizations have
effects that are far from negligible, and that will ultimately force
us to rethink our understanding of the results previously described
in this section.

\section{The trouble with semiconductor billiards - Soft walls and
small-angle scattering}

\subsection{Soft-walls and mixed phase-spaces}

The complexities involved in the theoretical calculations described
above typically forced semiconductor billiards to be stripped down
to their most basic elements for modeling (particularly in the early
90s, where computational capabilities were not what they are today).
The early models of dynamics in semiconductor microstructures
assumed hard-walls (i.e., the location of the wall in terms of
particle reflection is independent of energy) and geometries where
the phase space is either completely chaotic (ergodic) or completely
regular. However, semiconductor billiards are defined by
electrostatic depletion of selected regions of the 2DEG, using
either negatively biased surface-gates or by locally etching away
the dopants. This naturally results in a soft-wall
potential~\cite{LauxSurfSci88, KumarPRB90}, often best approximated
as parabolic~\cite{BerggrenChaos96}. This profile results
generically in a mixed phase-space, containing both chaotic and
regular regions; the ramifications of this were explored by
Ketzmerick~\cite{KetzmerickPRB96}. Trapping by the infinite
hierarchy of Cantori at the boundaries between regular and chaotic
regions in the phase space causes the escape time probability
distribution to take a power-law form $P(t) \sim t^{-\beta}$ with
exponent $\beta \leq 2$. This was expected via semiclassical
arguments to produce a power-law distribution of enclosed areas
$N(\theta) \sim \theta^{-\gamma}$ for large $\theta$ and
statistically self-similar fractal conductance fluctuations with
fractal dimension $D_{F} = 2 - \gamma/2$ for $\gamma \leq 2$ (and
$D_{F} = 1$ for $\gamma > 2$)~\cite{KetzmerickPRB96}.

\begin{figure}
\includegraphics[width=\linewidth]{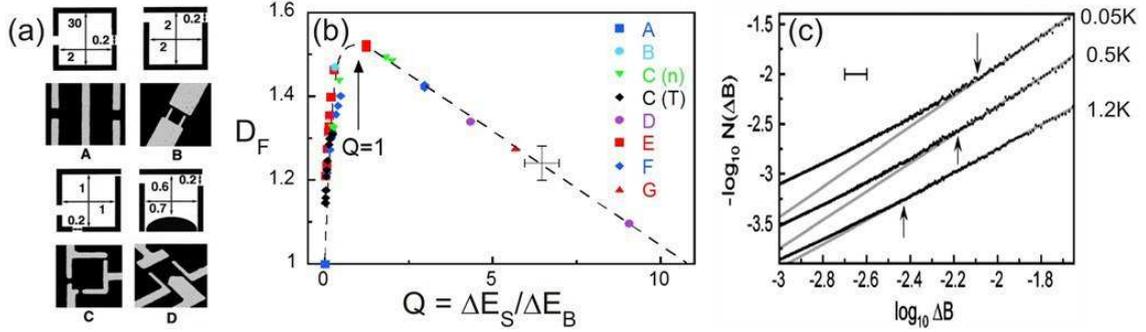}
\caption{(a) Gate schematics (top) and scanning electron micrographs
(bottom) for the seven billiard devices studied in
Ref.~\cite{MicolichPRL01}. Devices E, F and G are identical to B
with with widths of $1$, $0.6$ and $0.4~\mu$m, respectively. The
digits in (a) indicate lengths/widths in $\mu$m; Ti/Au gates appear
grey on the GaAs surface (black). (b) Measured fractal dimension
$D_{F}$ versus $Q$, which quantifies the resolution of quantized
energy levels within the billiard~\cite{MicolichPRL01}. (c) Fractal
scaling plot for $G(B)$ data from the `disrupted' billiard from
Ref.~\cite{MicolichPRB04} at $T = 50$~mK (top), $500$~mK and $1.2$~K
(bottom). The arrows indicate the lower cutoff for fractal scaling.}
\end{figure}

The observation of statistically self-similar fractal conductance
fluctuations in billiards was independently reported first by
Micolich {\it et al}~\cite{MicolichJPCM98} and soon thereafter by
Sachrajda {\it et al}~\cite{SachrajdaPRL98}.\footnote{n.b., Exact
self-similarity in a semiconductor Sinai billiard was reported by
Taylor {\it et al}~\cite{TaylorPRL97} a year prior.} In both cases,
fractal dimensions $1 < D_{F} < 2$ were reported, with preliminary
analyses reporting dependencies of $D_{F}$ on external parameters
such as gate voltage and temperature~\cite{MicolichJPCM98,
SachrajdaPRL98}.

A subsequent, more extensive study of the dependence of $D_{F}$ on a
range of parameters involving seven separate devices (Fig.~4(a)) by
Micolich {\it et al}~\cite{MicolichPRL01} lead to a remarkable
finding: a plot of $D_{F}$ versus a single parameter $Q$ for many
$G(B)$ traces all fall onto a single curve (Fig.~4(b)), despite
being obtained from seven different sizes and shapes of device with
differing number of 1D modes $n$ passing through the QPC entrance
and exit ports, electron mean free path $\ell$ and temperature $T$.
The parameter $Q$ is the ratio of the average energy level spacing
$\Delta E_{S}$ to the average energy broadening $\Delta E_{B}$ for
the dot, taking the form:

\begin{equation}
  Q = \frac{\Delta E_{S}}{\Delta E_{B}} = \frac{2\pi\hbar^{2}}{m^{*}A} \frac{1}{\sqrt{(\hbar/\tau_{q})^{2}+(k_{B}T)^{2}}}
\end{equation}

\noindent where $m^{*}$ is the electron effective mass and
$\tau_{q}$ is the lifetime of the quantum states, obtained using a
skipping orbit analysis~\cite{BirdSurfSci96, BirdPSSB97,
BirdJPCM98}.\footnote{This is ultimately an unentangleable
combination of the phase coherence time $\tau_{\phi}$ and the escape
time related to a skipping orbit finding a QPC through which to exit
the dot.} Remarkably, $D_{F}$ is found to take its maximum value
$\sim 1.52$ at $Q = 1$, with $D_{F}$ falling in either direction,
reaching the non-fractal value $D_{F} = 1$ in the limits $Q = 0$ and
$Q \sim 10$. The trend in the small $Q$ limit is particularly
interesting; small $Q$ is typically achieved via the denominator
through increases in $T$ and/or decreases in $\tau_{q}$. Based
purely on Ref.~\cite{KetzmerickPRB96}, one would expect that a
reduction in the coherence length (and thus $\tau_{q}$) independent
of other billiard parameters would reduce the scaling range over
which fractal behaviour was observed but not alter $D_{F}$ itself.
Indeed, later work (Fig.~4(c)) showed that the scaling range instead
increases with increasing temperature concurrent with a falling
$D_{F}$~\cite{MicolichPRB04}. Calculations by Budiyono and
Nakamura~\cite{BudiyonoCSF03} based on a semiclassical Kubo
formalism obtained a $D_{F}$ that decreases smoothly with increasing
$T$; an interesting side-aspect of these calculations is a
connection to the Weierstrass function model~\cite{TaylorPRB98,
MicolichEPL00} for the exact self-similarity reported in Sinai
billiards~\cite{TaylorPRL97}. Hennig {\it et al}~\cite{HennigPRE07}
provide an interesting alternative, suggesting that the underlying
classical dynamics also lead to fractal conductance fluctuations. As
such, $G(B)$ is a superposition of two fractals, one due to quantum
interference, which is suppressed by decoherence to reveal fractal
fluctuations due to the classical processes. Note that the data
below the lower cut-off (i.e., low $\Delta B$ limit) in Fig.~4(c) is
linear with a different slope. This might be indicative of a process
similar to that proposed by Hennig {\it et al}, however, care is
needed as this data is close to the resolution limit/noise floor;
further experiments in this direction are warranted. Several studies
suggest that the semiclassical theory proposed by
Ketzmerick~\cite{KetzmerickPRB96} is only part of a larger picture.
A complete discussion is beyond the scope of this paper, but most
significantly, mixed phase space does not appear essential to the
generation of fractal conductance fluctuations~\cite{TakagakiPRB00,
LouisPRB00}, which are also expected theoretically for completely
chaotic~\cite{LouisPRB00, BenentiPRL01, PinheiroBJP06} and
completely integrable~\cite{GuarneriPRE01} billiards.

Considering Ketzmerick's theory, an interesting parameter to vary is
the softness of the billiard walls. The first studies were performed
by using a pair of billiards defined, one above the other, in a
double quantum well heterostructure to provide identical geometries
with different wall softness~\cite{MicolichAPL02}. Self-consistent
potential simulations~\cite{FromholdPhysB98} showed that the profile
gradient at the Fermi energy differed by a factor of three between
the $90$ and $140$~nm deep billiards~\cite{MicolichAPL02}. The data
in both billiards was consistent with Eq.~5; the only difference
compared to the data in Fig.~4(b), which contains the shallower
billiard data as device C, is that the deeper billiard (harder
potential) gave a lower maximum $D_{F} \sim 1.45$ at $Q =
1$~\cite{MicolichPRB04}. This difference in maximum $D_{F}$ is
small, and may be due to competing effects given the complexities of
this device~\cite{MicolichAPL02}.

The effect of soft-wall profile was also studied by Marlow {\it et
al}~\cite{MarlowPRB06}. A further 21 devices were investigated: $11$
Ga$_{0.25}$In$_{0.75}$As/InP billiards and $5$
Ga$_{0.25}$In$_{0.75}$As/InAlAs billiards of various geometries, and
$5$ GaAs quantum wires. As demonstrated by Martin {\it et
al}~\cite{MartinSM03}, the potential profile for the etched
GaInAs/InP billiards\footnote{The substitution of InAlAs for InP
should not affect the profile of the billiard walls. The slight
change in band-gap (from $\sim 1.42$~eV to $\sim 1.55$~eV) should
not significantly affect the 2DEG confinement either. There is a
large difference in disorder however, as described in the text and
Ref.~\cite{MarlowPRB06}.} is ten times steeper than for the surface
gated AlGaAs/GaAs billiards in Ref.~\cite{MicolichPRL01,
MicolichPRB04}. Despite this, the data for all of these devices
closely overlays that in Fig.~4(b), suggesting that wall profile is
not a determining factor for $D_{F}$. The data obtained for
quasi-ballistic and diffusive GaAs quantum wires was also fractal
and fit on the curve in Fig.~4(b). This is consistent with earlier
work showing fractal conductance fluctuations in quasi-ballistic
gold nanowires~\cite{HeggerPRL96}, and predictions that fractal
conductance fluctuations survive diffusive
transport~\cite{PinheiroBJP06}.

Ultimately, the results obtained by Micolich {\it et
al}~\cite{MicolichPRL01, MicolichPRB04} and Marlow {\it et
al}~\cite{MarlowPRB06} suggest that the statistical properties of
the conductance fluctuations, at least in terms of fractal
dimension, are independent of the geometry of the billiard
walls\footnote{Note that other statistical measures of the
fluctuations, such as the power spectrum, are also linked to fractal
dimension~\cite{BarnsleyBook88}.}. This is unexpected given that
transport in these devices is nominally `ballistic'. Instead, the
statistical properties via Eq.~5 and $Q$, depend only on $A$, $T$
and $\tau_{q}$ irrespective of the specifics of the sample itself.
This `universality' bears a strong resemblance to that of Universal
Conductance Fluctuations (UCF) in mesoscopic metal films and MOSFETs
(see, for example, Fig.~3 of Skocpol~\cite{SkocpolPS87}) where the
transport is diffusive rather than ballistic. This raises
interesting and radical questions: Even though $\ell >> L,W$ for
billiards and they are ballistic by traditional
measures~\cite{BeenakkerSSP91}, is the transport really ballistic?
Or is it diffusive, such that a form of scattering essentially
undetected by the mobility, dominates the transport?

\subsection{Remote ionized dopants and small-angle scattering}

The 2DEG in an AlGaAs/GaAs heterostructure is normally populated by
the ionization of remote Si dopants. These positively charged
ionized dopants interact electrostatically with the electrons in the
2DEG causing significant scattering. High mobility 2DEGs are
obtained through a process known as `modulation
doping'~\cite{DingleAPL78}, where the Si dopants are separated from
the 2DEG by a $20 - 100$~nm undoped AlGaAs spacer layer. The key to
modulation doping is that ionized dopants are converted from
large-angle scattering sites to small-angle scattering sites by
increases in the dopant-2DEG separation. This effectively `hides'
some of the ionized dopant scattering because the mobility is
weighted towards large-angle scattering~\cite{ColeridgePRB91,
MacLeodPRB09} (see $\S5$). Nonetheless, the 2DEG still feels the
effect of the ionized dopants, which present as a low-level ($\sim 1
- 10$~meV) random `disorder potential' for the
electrons~\cite{NixonPRB90}. The length scale of this scattering is
set by the 2DEG-donor separation~\cite{JuraNP07}, usually of order
$20 - 100$~nm, is much smaller than the typical billiard width
($\sim 0.6 - 2~\mu$m) and the large-angle scattering length $\ell
\sim 2 - 20~\mu$m.

At the time of the initial studies~\cite{JalabertPRL90, MarcusPRL92,
BarangerPRL93, ChangPRL94, BeenakkerSSP91}, knowledge of the effect
that the underlying disorder potential had on ballistic transport in
microstructures was relatively low, although numerical simulations
showed it could significantly affect transport at length scales much
smaller than $\ell$~\cite{NixonPRB91}. The first focussed
consideration in terms of billiards was by Lin, Delos and
Jensen~\cite{LinChaos93}, and involved adding a random angle $[-\pi,
+\pi]$ at each wall reflection in a semiclassical dynamical model
for stadium and circle billiards. The focus was on the statistics,
and while the disorder in this model had little effect for the
stadium (as expected given the ergodic dynamics), it destroyed the
power-law tail in the area distribution for the circular billiard.
This gave an exponential distribution, as expected for the stadium
billiard, albeit with a different exponent~\cite{LinChaos93}. This
should produce a Lorentzian $R(B = 0)$ peak for both stadium and
circle billiards. Chang {\it et al}~\cite{ChangPRL94} approached the
same problem differently, adding a random energy
$[-W_{dis}/2,W_{dis}/2]$ at every fifth lattice site and linearly
interpolating in between in quantum calculations on a discretized
lattice~\cite{BarangerPRB90}. They concluded that the linear $R(B =
0)$ peak for the circular billiard should not be destroyed by a
disorder potential with a strength chosen to match experimental
conditions, with roughness in the dot wall being more
crucial~\cite{ChangPRL94}. Ji and Berggren~\cite{JiPRB95,
BerggrenChaos96} made the first investigation at the single $G(B)$
trace level (as opposed to the statistics of area distributions or
averaged $R(B = 0)$ peak lineshape) for stadium-shaped billiards
using a discrete lattice model~\cite{WangPRB94}. Their model showed
that changes in disorder potential can have a significant impact on
$G(B)$ as a magnetofingerprint for the electron
dynamics~\cite{JiPRB95}, as predicted for UCF by Feng, Lee and
Stone~\cite{FengPRL86}, and demonstrated for diffusive quantum wires
by Taylor {\it et al}~\cite{TaylorCJP92} and Klepper {\it et
al}~\cite{KlepperPRB91}.

\begin{figure}[b]
\includegraphics[width=\linewidth]{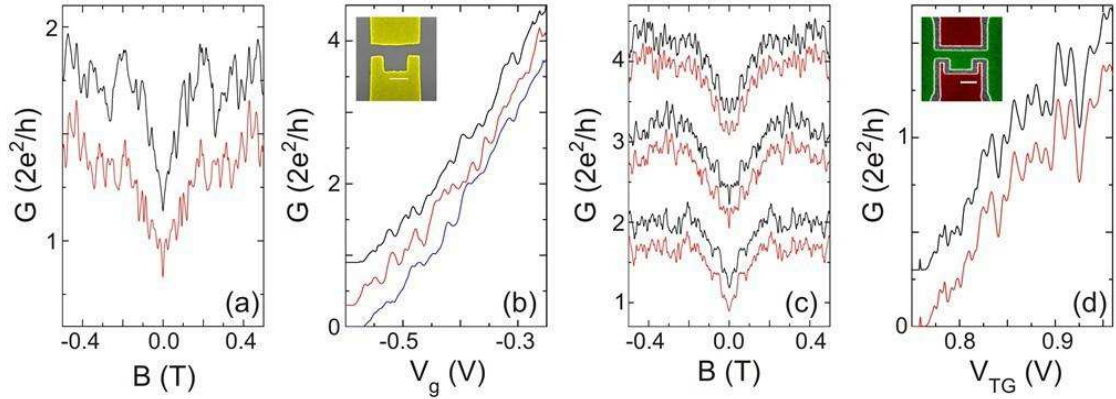}
\caption{Conductance $G$ versus magnetic field $B$ and gate voltage
$V$ from (a/b) a modulation-doped billiard and (c/d) an undoped
billiard. False colour scanning electron micrographs of these
devices are shown inset to (b) and (d) respectively. The scale bars
indicate $500$~nm. The two devices have $\ell = 2.7$ and $2.1~\mu$m
respectively. In (b), the gate bias $V_{g}$ is applied to both metal
gates (yellow). In (d), the gate bias $V_{TG}$ is applied to the
n$^{+}$-GaAs top-gate (green), with the side-gates (red) grounded.
In both cases $B = 0$. The traces in (a) were obtained at $V_{g} =
-430$ (top) and $-451$~mV (bottom); these voltages were chosen to
match $G(B = 0)$ before and after cycling. $G(B)$ traces obtained
with common $V_{g}$ do not match either (see Ref.~\cite{SeePRL12}).
The three pairs of traces in (c) were obtained at $V_{TG} = +955$
(top), $+945$ (middle) and $+935$~mV (bottom). The traces are shown
in each case from top down: first (black), second (red) and third
(blue) cooldown. Traces are vertically offset for clarity by: (a)
bottom trace $-0.3$; (b) top trace $+0.9$, middle trace $+0.3$; (c)
from bottom by $-0.3$, $0$, $+0.7$, $+1.0$, $+1.7$, $+2.0$; (d) top
trace $+0.3$; all in units of $2e^{2}/h$.}
\end{figure}

Recently, a technique known as scanning gate microscopy
(SGM)~\cite{ErikssonAPL96} has given remarkable insight into the
`disorder potential' and its influence on electron
transport~\cite{TopinkaNat01, JuraNP07, AidalaNP07}. SGM studies of
2DEGs show significant branching of electron
flow~\cite{TopinkaNat01}, and even in samples with very long mean
free paths ($\ell > 10~\mu$m) significant deviations in propagation
persist at scales $\sim 100$~nm~\cite{JuraNP07}. An SGM study of a
ballistic focussing structure by Aidala {\it et
al}~\cite{AidalaNP07} gives a remarkable visual demonstration of the
impact that small-angle scattering has on even quite short and
simple ballistic transport paths between QPCs. Although small-angle
scattering clearly doesn't prevent ballistic focussing from being
observed electrically~\cite{vanHoutenPRB89}, it certainly forces one
to question whether the concept of semiconductor billiards with long
straight-line trajectories bouncing specularly off the billiard
walls (see e.g., Fig.~2 of Ref.~\cite{JensenNat95} or Fig.~1/6 of
Ref.~\cite{HellerPT93}) is a realistic picture? It also raises the
question of the extent to which small-angle scattering governs the
dynamics and resulting electrical properties of these devices.

Scannell {\it et al}~\cite{ScannellArXiv11} and See {\it et
al}~\cite{SeePRL12} have recently performed experiments aimed at
illuminating the relative importance of disorder and boundary
scattering in semiconductor billiards. These experiments rely on an
important feature of the dopant physics of AlGaAs/GaAs
heterostructures and the newly developed capability to make
billiards where the 2DEG is populated electrostatically using a gate
enabling the dopants to be removed~\cite{SeeAPL10} and the effect of
small-angle dopant scattering on transport to be clearly observed.
Si dopants in modulation-doped Al$_{x}$Ga$_{1-x}$As/GaAs
heterostructures with $x > 0.22$ can take one of two possible
configurations~\cite{LongSST93, BuksPRB94}. The first is as a
shallow hydrogenic donor with a positive or neutral charge
($d^{+}$/$d^{0}$) sitting at a substitutional site. The second,
known as a DX center, involves a lattice distortion that pulls the
Si donor into an interstitial position along the $\langle111\rangle$
direction, producing a deep trap. At temperatures below $\sim
150$~K, the DX centers normally bind two electrons and have a
negative charge (DX$^{-}$), but can be excited thermally or
optically to detrap into DX$^{0}$ and DX$^{+}$ states. As described
earlier, the ionized dopants impose a `disorder potential' on the
2DEG~\cite{NixonPRB90}, with a density of scattering sites $N =
N(d^{+}) + N($DX$^{-}) + N($DX$^{+})$, where $N(...)$ is the density
of a particular dopant configuration. While $N$ is normally fixed
for a given heterostructure and gate-voltage configuration during
cooldown~\cite{BuksPRB94}, the allocation of particular Si site as
$d$ or DX varies varies each cooldown. Thus the resulting disorder
potential is fixed for a single cooldown, leading to the
repeatability of $G(B)$ traces shown in Fig.~3(b) and
Ref.~\cite{MarcusChaos93}. In contrast, the disorder potential
should change markedly on separate cooldowns, and this would produce
one of two results. If the transport in the billiard is truly
ballistic, such that the disorder is negligible and the electron
trajectories are determined only by scattering from the boundaries,
then $G(B)$ should remain the same in subsequent cooldowns
(providing $T > 150$~K in between). Alternatively, if the disorder
potential does affect transport, then we should see some level of
change in $G(B)$, as predicted by Feng, Lee and
Stone~\cite{FengPRL86}.

Figure~5(a) shows $G(B)$ traces obtained on separate cooldowns of
the modulation-doped device inset to Fig.~5(b). There is clearly a
striking difference between the traces indicative of the disorder
potential affecting the transport. The gate voltage was adjusted
slightly on the second cooldown in order to match $G(B = 0)$ and
obtain similar injection conditions~\cite{AkisJJAP97, BirdCSF97},
but See {\it et al} observe the same behaviour if $V_{g}$ is kept
the same, or takes other values as part of an extensive search for
matching $G(B)$ traces~\cite{SeePRL12}. As Fig.~5(b) shows, the
change in disorder potential upon thermal cycling also alters the
gate characteristics; the fluctuations there also originate from
quantum interference~\cite{BirdPRL99}. This thermally-induced
irreproducibility is also evident in a parallel study by Scannell
{\it et al}~\cite{ScannellArXiv11}, where the device is cooled to
$300$~mK and $G(B)$ is measured, the temperature is raised to an
intermediate temperature $T_{1}$ for 30 minutes and returned to
$300$~mK, $G(B)$ is measured again, the temperature is raised to
$T_{2} > T_{1}$, re-cooled and measured again, and so on until
$T_{n} = 298$~K. The various post-cycling $G(B)$ traces are compared
to the first cooldown $G(B)$ trace using a cross-correlation
analysis~\cite{TaylorPRB97}, with a precipitous drop in the
correlation coefficient $F$ from $1$ to $0$ obtained at $T \sim 100
- 175$~K for AlGaAs/GaAs and GaInAs/InP billiards and GaAs quantum
wires. This temperature dependence points strongly to ionized dopant
scattering playing a major role in influencing transport in these
devices.

The `acid test' for the hypothesis that the disorder potential
dominates transport is to remove the dopants entirely. This requires
a complete redesign of the device; a metallic gate needs to be added
and this needs to overlap the contacts to the 2DEG to ensure
electrical continuity. This is achieved using a degenerately-doped
GaAs cap layer as the gate, which enables self-alignment of the gate
to the contacts~\cite{KaneAPL93, SeeAPL10}. The n$^{+}$-GaAs cap is
divided into three gates by shallow wet etching as shown inset to
Fig.~5(d); the central gate (green) is biased positively to $V_{TG}$
to electrostatically populate the 2DEG, the side-gates (red) can be
used to tune the device further, but remain grounded here.
Figure~5(c) shows $G(B)$ before and after thermal cycling to $300$~K
for three different $V_{TG}$, in each case the MCF are reproducible
with high fidelity. As Fig.~5(d) shows, the gate characteristics are
also highly reproducible.

Given the MCF's role as a `magnetofingerprint'~\cite{FengPRL86} of
electron transport within the dot, Fig.~5 demonstrates that the
disorder potential has a significant, perhaps even dominant,
influence on the electron dynamics. We now discuss an explanation
for how and why this occurs, the implications regarding past
results, and some thoughts on follow-up experiments to further
elucidate the role that disorder plays in billiards.

\section{Discussion}

Historically, a major focus in mesoscopic physics has been to
improve the mobility to access new physical regimes that are
otherwise obscured by disorder. This was a key driver for work on
the fractional quantum Hall effect~\cite{StormerRMP99, PanPRL11} and
ballistic transport devices (e.g., quantum wires, dots and
billiards.) as per Figs.~2(a-c). It is vital to note that the
mobility, and by association $\ell$, are only part of the picture.
The mobility is a measure of momentum decay and it contains a
weighting term $1~-~$cos$~\theta$~\cite{ColeridgePRB91}, where
$\theta$ is the scattering angle, that makes it preferentially
sensitive to large-angle scattering events~\cite{ColeridgePRB91,
MacLeodPRB09, JuraNP07} Figures~2(d/e) present a more refined
picture for the diffusive and ballistic regimes; here large-angle
scattering sites are black and small-angle scattering sites are grey
(darker shades represent a propensity towards larger scattering
angles). The two pictures of diffusive transport (a/d) are
essentially the same, and in either case, it is impurity scattering
that dominates over boundary scattering. The two pictures for
ballistic transport (b/e) are very different; as in $\S4.2$ the
ionized dopants present as small-angle scatterers with a scattering
length set by the spacer thickness~\cite{JuraNP07}, which is much
smaller than $L$, $W$ and $\ell$. These deflect the electron
trajectories between the walls as shown, suggesting a more complex
picture with a less clear-cut distinction between the traditional
diffusive and ballistic regimes. Bearing in mind the thermal
redistribution in dopant potential~\cite{ScannellArXiv11, SeePRL12},
this explains why the MCF changes so much between cool-downs and
differs when studying the same geometry on two separate chips (see
Fig.~3 of Ref.~\cite{MarcusPRL92}).

From this perspective, the MCF in modulation-doped billiards is
probably best viewed as analogous to UCF in metal films and narrow
MOSFETs~\cite{LeePRL85, AltshulerJETP85, LiciniPRL85, KaplanPRL86,
SkocpolPRL86, LeePRB87, SkocpolPS87}. In both cases the transport is
dominated by disorder scattering, the fluctuations are reproducible
in a single sample/disorder configuration but vary between samples
and disorder configurations, are quenched by an increase in
temperature, and have certain universal statistical properties
independent of the sample shape and precise dopant
configuration~\cite{LeePRB87} -- for the UCF this is the amplitude
$\delta G$ which depends on the ratio of length $L$ to coherence
length $L_{\phi}$~\cite{SkocpolPS87}, whilst for MCF in billiards,
this appears to be the fractal dimension $D_{F}$ which depends on a
more complex combination of area $A$ and quantum lifetime
$\tau_{q}$, at fixed temperature~\cite{MicolichPRL01}. The billiard
walls should help rather than hinder this analogy; their action is
to repeatedly feed the electron back into the diffusive environment
inside the billiard~\cite{MarlowPRB06}. In a sense, the walls impose
reflective periodic boundary conditions, effectively converting the
billiard into a larger area, diffusive 2D system. The natural
correlations between these mirrored sub-units, combined with the two
QPCs acting as points of focus for the interference contribution may
explain the loss of amplitude universality, and its replacement by
the more complex universality in Fig.~4(b)~\cite{MicolichPRL01,
MicolichPRB04, MarlowPRB06}. Modeling a small-area, small-angle
diffusive system with reflective boundary conditions on the edge for
all but two points where trajectories can enter/exit and contribute
to $G(B)$, and comparing with traditional UCF~\cite{LeePRB87} may
give interesting insights. Alternatively, it may be interesting to
consider existing UCF models, but with the uniform random disorder
replaced by repeating small `tiles' of a common disorder
configuration.

Nonetheless, `ballistic' transport, where the walls rather than
disorder is the shaping force for a given trajectory, cannot be
neglected entirely in billiards. Comparing Figs.~2(b,c,e) it is
clear that the diffusive effect of small-angle scattering should
increase with trajectory length. As a result, the long trajectories
crucial to discriminating between chaotic and regular geometries in
semiclassical theories~\cite{JalabertPRL90, BarangerPRL93} are
likely destroyed by disorder such that all billiards are essentially
chaotic, irrespective of geometry, from the perspective of the tails
in the trajectory area distributions $N(\theta)$. As such, the
disorder has the same `chaos-inducing' effect as a Sinai
diffuser~\cite{SinaiRMS70, DoronPRL90, FromholdPhysB98}, but with a
scale and quantity more similar to antidot
arrays~\cite{YevtushenkoPRL00, DornPRB05} where chaotic transport,
weak localization and quantum interference fluctuations are also
observed. Shorter trajectories may well survive, and we see two
separate categories of these. The first are skipping orbits as
highlighted by Christensson {\it et al}~\cite{ChristenssonPRB98},
and used for studying phase-breaking in dots by Bird {\it et
al}~\cite{BirdSurfSci96, BirdPSSB97, BirdJPCM98}. These are likely
protected from disorder scattering by the short distance between
bounces combined with a field-induced curvature mechanism similar to
that involved in the quantum Hall effect~\cite{ButtikerPRB88}. The
feasibility of the survival of skipping orbits amidst the disorder
potential is evident in the SGM studies by Aidala {\it et
al}~\cite{AidalaNP07}. The survival of short ballistic trajectories
may also explain measurements of the symmetry of MCF in the
non-linear voltage regime as a function of the symmetry of billiard
shape~\cite{LofgrenPRL04, MarlowPRL06}. Combined, the work above
provides strong evidence that billiard shape is still exerts an
influence on the MCF.

The second category is periodic orbits, the classic example being
the diamond scar in square dots~\cite{AkisJJAP97, BirdCSF97}; this
also clearly survives small-angle scattering as observed in SGM
studies by Burke {\it et al}~\cite{BurkePRL10}, but scars with other
shapes have also been reported very recently~\cite{AokiPRL12}. It is
easy to consider this as a straightforward case of wavefunction
scarring~\cite{HellerPRL84, WilkinsonNat96, AkisJJAP97}, but there
may be more to it. The diffusive effect of disorder should not only
affect the unstable periodic orbit underlying the wavefunction scar,
but it will do so differently between devices, between different
cooldowns of the same device, and possibly also as parameters such
as the gate voltage are changed. The latter may partially explain
the observation of some but not all of the expected scar states in
previous experiments on modulation-doped dots~\cite{BirdPRL99}.
Wavefunction scarring in dots has recently been linked to a theory
known as `Quantum Darwinism'~\cite{ZurekNP09}, where preferred
states within the dot survive coupling with the environment, which
acts as the `selection pressure' in the theory~\cite{BrunnerPRL08}.
From an experimental perspective~\cite{BirdPRL99}, it may be that
disorder presents an additional selection pressure, acting in
partnership with environmental coupling, that eliminates scar states
connected to much more complex underlying periodic orbits. From this
perspective, it would be interesting to extend recent SGM studies of
dots~\cite{BurkePRL10, AokiPRL12} to undoped heterostructure system
to see if a greater variety of scars are observed. It would also be
interesting to perform SGM studies of undoped 2DEGs and ballistic
focussing structures more generally to establish the level of
improvement provided by eliminating the ionized dopants. The current
architecture, with a gate above the 2DEG, is unsuitable for SGM; an
alternative would be to use an inverted structure~\cite{MeiravAPL88,
MeiravPRL90}, possibly using deep etching instead of surface gates
and/or patterning of the doped gate layer using focussed ion-beam
lithography, as in the bilayer billiard~\cite{MicolichAPL02}.

We conclude by attempting to `close the loop' and reconsider
ballistic microstructures as devices for studying dynamical chaos in
the quantum limit. The data in Fig.~5(c/d) shows that removal of the
modulation doping instils thermal robustness in the MCF obtained
from the device. Correspondingly, one would expect a significant
reduction in small-angle scattering, despite this, substantial
disorder will remain. This includes background impurities, some
ionizable, some not, as well as interface roughness, defects in the
walls of the device, etc. We suspect that this remnant disorder will
prevent perfectly identical $G(B)$ from being obtained in separate
undoped devices with nominally identical geometry, obviating the
truly ballistic quantum dots envisioned
theoretically~\cite{JalabertPRL90, BarangerPRL93, JensenNat95}. One
path forward might be the approach used by Chang {\it et
al}~\cite{ChangPRL94}, where a $6 \times 8$ array of nominally
identical dots was studied to average out the MCF. This averaging of
the MCF essentially constitutes an averaging of the dot disorder
potentials to expose the effect of the common lithographic geometry.
Attempting this using arrays of undoped dots in clean
heterostructures with low background impurity densities may bring
these devices close to a true `ballistic' limit in terms of
semiclassical dynamics, providing of course that the lithography
defining their walls can be made sufficiently identical. This would
be a useful challenge for future work.

Ultimately, while single quantum dots are not sufficiently
`ballistic' for use in studies of quantum chaos in our opinion, they
have been shown to be sufficiently ballistic for potential
applications~\cite{GoodnickIEEE03} as ballistic
rectifiers~\cite{SongAPL01} and Y-branch
switches~\cite{WorschechAPL01}. The reduced small-angle scattering
in undoped quantum dots offers much potential in this regard, their
thermal robustness suggests they may also offer a path to devices
with significantly reduced charge noise, which is a well known
effect of fluctuations in the charge state of ionized
dopants~\cite{BuizertPRL08}.

\section{Acknowledgements}

We thank the following for helpful discussions, contributions and
support to this research over many years: M.~Aagesen, R.~Akis,
N.~Aoki, Y.~Aoyagi, V.~Bayot, J.P.~Bird, S.~Bollaert, S.A.~Brown,
A.M.~Burke, A.~Cappy, J.~Cooper, A.G.~Davies, C.P.~Dettmann,
L.~Eaves, S.~Faniel, M.S.~Fairbanks, I.~Farrer, D.K.~Ferry,
T.M.~Fromhold, C.~Gustin, B.~Hackens, K.~Ishibashi, O.~Klochan,
P.E.~Lindelof, E.H.~Linfield, L.D.~Macks, P.K.~Morse, G.P.~Morriss,
R.~Newbury, Y.~Ochiai, S.M.~Reimann, D.A.~Ritchie, L.~Samuelson,
I.~Shorubalko, T.~Sugano, W.R.~Tribe and X.~Wallart.

\end{document}